\documentclass{aastex}
\newcommand{\ud}{\,\mathrm{d}}

\usepackage{spr-astr-addons}
\usepackage{url}

\begin{document}

\title{The Generalized Uncertainty Principle and the Friedmann equations}
\slugcomment{manuscript submitted to \textit{Astrophysics and Space Science}}
\shorttitle{The Generalized Uncertainty Principle and the Friedmann equations}
\shortauthors{Barun Majumder}

\author{Barun Majumder} 
\affil{Department of Physical Sciences,\\Indian Institute of Science Education and Research (Kolkata),\\
Mohanpur, Nadia, West Bengal, Pin 741252~, India. \\ email: barunbasanta@iiserkol.ac.in}

\begin{abstract} 
The Generalized Uncertainty Principle (or GUP) affects the dynamics in Plank scale. So the known equations of physics are expected to
get modified at that very high energy regime. Very recently authors in \cite{z6} proposed a new Generalized Uncertainty Principle (or GUP) with
a linear term in Plank length. In this article, the proposed GUP is expressed in a more general form and the effect is studied for the
modification of the Friedmann equations of the FRW universe. In the midway the known entropy-area relation get some new correction terms, the leading
order term being proportional to $\sqrt{\mathrm{Area}}$\,.
\end{abstract}
\vspace*{8mm}

If we apply Clausius relation of thermodynamics ($\delta Q=T \ud S$) to the apparent horizon of any finite dimensional FRW universe with any
spatial curvature we get the Friedmann equations \cite{a4}. The entropy is a quarter of the apparent horizon area and the temperature
of the apparent horizon has the form $T= \frac{1}{2 \pi \tilde{r}_A}$ are the accompanying two considerations for the derivation of the Friedmann
equations. Here $T$ is the temperature of the apparent horizon with radius $\tilde{r}_A$. $\delta Q$ denotes the amount of energy crossing the
apparent horizon in an infinitesimal time interval. Using the tunnelling approach method as proposed by Parikh and Wilczek \cite{a14c} we can also
prove \cite{a15} that the apparent horizon of a FRW universe in any finite dimension has an associated Hawking temperature given by
$T= \frac{1}{2 \pi \tilde{r}_A}$. The application of the first law of thermodynamics in other gravity theories such as Gauss-Bonnet gravity, Lovelock
gravity and also in various braneworld scenarios can be found in \cite{a5},\cite{a6},\cite{a}. Recently it was also discussed that the acceleration is due to an entropic force naturally arising from the information storage on the horizon surface screen \cite{req1}.

The idea that the uncertainty principle could be affected by gravity was first given by Mead \cite{z1}. Later modified commutation relations
between position and momenta commonly known as Generalized Uncertainty Principle ( or GUP ) were given by candidate theories of quantum gravity ( String Theory,
Doubly Special Relativity ( or DSR ) Theory and Black Hole Physics ) with the prediction of a minimum measurable length \cite{a16},\cite{m3},\cite{m4},\cite{z4}. Similar
kind of commutation relation can also be found in the context of Polymer Quantization in terms of Polymer Mass Scale \cite{z5}.
GUP addresses the existence of a minimal length scale (generally the Plank length scale is considered). We can use GUP to modify the black
hole thermodynamics and thereby we can get correction terms for the Bekenstein-Hawking entropy \cite{a21},\cite{a22}. Using GUP we can also modify the
Hawking temperature which eventually prevents the total evaporation of a radiating black hole. We get an inert remnant with zero entropy and finite
temperature in the final stages of evaporation \cite{a19},\cite{a20},\cite{bh112}. For the FRW universe in any finite ($n+1$)-dimension, we can modify the Friedmann equations \cite{a13},\cite{a} with the help of the first law of thermodynamics on the apparent horizon. The modification is due to GUP because it affects
the entropy-area relation by adding corrections.

The authors in \cite{z6} proposed a GUP which is consistent with DSR theory, String theory and black hole physics and which says
\begin{equation}
\left[x_i,x_j\right] = \left[p_i,p_j\right] = 0 ,
\end{equation}
\begin{eqnarray}
\label{e2}
[x_i, p_j] =&& i \hbar \bigg[  \delta_{ij} -  l  \left( p \delta_{ij} + \frac{p_i p_j}{p} \right) \nonumber \\
&&+ l^2  \left( p^2 \delta_{ij}  + 3 p_{i} p_{j} \right)\bigg]\,, 
\end{eqnarray}
and
\begin{equation}
\label{ee1}
 \Delta x \Delta p \geq \frac{\hbar}{2} \left[ 1 - 2 l \langle p \rangle + 4 l^2 \langle p^2 \rangle \right]~~,
\end{equation}
where $ l=\frac{a_0 l_{p}}{\hbar} $. Here $ l_{p} $ is the Plank length ($ \approx 10^{-35} m $) and it is generally assumed that $ a_0 = 1 $. It is evident that this new physical length scale which is of the order of $a_0 l_{p} $ cannot exceed the electroweak length scale $ \sim 10^{17} l_{p}$ which surely implies $a_0 \leq 10^{17}$. The above equations are approximately covariant under DSR transformations but not Lorentz covariant \cite{z4}. These equations also imply
\begin{equation}
\delta x \geq \left(\delta x \right)_{min} \approx a_0l_{p}
\end{equation}
and
\begin{equation}
\delta p \leq \left(\delta p \right)_{max} \approx \frac{M_pc}{a_0}
\end{equation}
where $ M_p $ is the Plank mass and $c$ is the velocity of light in vacuum. The effect of this proposed GUP is well studied recently for some well known
physical systems in \cite{z6},\cite{z7},\cite{bar1}.

In this article we will reinvestigate the results of \cite{a13},\cite{a} using this newly proposed GUP \cite{z6}. We will apply this GUP for the
calculation of the modified entropy. Here we will make the assumption that the apparent horizon has the entropy which is the calculated modified entropy.
Then we will apply the Clausius relation of thermodynamics to the apparent horizon to get the newly modified Friedmann equations which governs
the dynamical evolution of the universe. If we now see equation (\ref{e2}) we can easily notice that the uncertainty relation has a linear term
in the Plank length which is absent in the earlier version of GUP \cite{a16}. Here we write the GUP of equation (\ref{e2}) in a more general
form as
\begin{equation}
\label{n1}
\delta x \delta p \geq \, 1 + \beta \,l_p \,\delta p + \alpha^2 \,l_p^2\, (\delta p^2)\,.
\end{equation}
Throughout the whole process we will consider $\hbar=G=c=k_B=1$, where $G$ is the Newton constant, $c$ the velocity of light in vacuum and $k_B$ the
Boltzmann constant. Here $l_p$ is the Plank length. $\beta$ is the coefficient of $l_p$ and $\alpha^2$ is the coefficient of $l_p^2$. $\beta$
essentially highlights the effect of the linear term in Plank length of the uncertainty relation. We keep the freedom in our hand to choose
the parameters $\alpha$ and $\beta$. For example $\beta=0$ gives back the earlier version of GUP as discussed in \cite{a16}. We can further
tune $\beta=-2a_0$ and $\alpha^2=4a_0^2$ so that we get back (\ref{ee1}). From equation (\ref{n1}) we can write the form of the momentum uncertainty as
\begin{eqnarray}
\label{u3}
\delta p &&\geq \frac{1}{\delta x} \Bigg[ \frac{(\delta x)^2}{2 \alpha^2 l_p^2} \bigg(1-\frac{\beta l_p}{\delta x}\bigg) \nonumber \\
&& \bigg\{1\pm \sqrt{1-
          \frac{4\alpha^2 l_p^2}{(\delta x)^2 \big(1-\tfrac{\beta l_p}{\delta x}\big)^2}}\bigg\}\Bigg] \nonumber \\
         &&\geq \frac{1}{\delta x} f_{GUP} (\delta x^2) \, ,      
\end{eqnarray}
where 
\begin{eqnarray}
f_{GUP} (\delta x^2) =&& \Bigg[ \frac{(\delta x)^2}{2 \alpha^2 l_p^2} \bigg(1-\frac{\beta l_p}{\delta x}\bigg)\nonumber \\
&&\bigg\{1\pm \sqrt{1-
\frac{4\alpha^2 l_p^2}{(\delta x)^2 \big(1-\tfrac{\beta l_p}{\delta x}\big)^2}}\bigg\}\Bigg]\,.
\end{eqnarray}
$f_{GUP}(\delta x^2)$ measures the amount of departure from our usual Heisenberg uncertainty principle. We consider the negative sign
for $f_{GUP}(\delta x^2)$ as the positive sign has no physical meaning. We now consider a ($n+1$)-dimensional FRW universe with line element
\begin{equation}
\ud s^2 = - \ud t^2 + a^2 \bigg(\frac{\ud r^2}{1-kr^2} + r^2 \ud \Omega_{n-1}^2\bigg) \, ,
\end{equation}
where $\ud \Omega_{n-1}^2$ is the line element of a ($n-1$)-dimensional unit radius sphere, $a$ is the scale factor and $k$ defines the
curvature of the spatial section. Without the whole evolution history of the universe we cannot say anything about the cosmological
event horizon. But a dynamical apparent horizon always exist in the FRW universe because it is a local quantity of spacetime.
The apparent horizon is a marginally trapped surface with vanishing expansion. The location of the apparent horizon is given by
\begin{equation}
\tilde{r}_A = \frac{1}{\sqrt{H^2 + \frac{k}{a^2}}} \, ,
\end{equation}
where $\tilde{r}_A$ is due to the re-definition $\tilde{r}_A = ar$. Here $H(=\frac{\dot{a}}{a})$ is the Hubble parameter. The apparent
horizon and the cosmological event horizon coincide only for $k=0$. If we assume that the apparent horizon is associated with entropy
$S=\frac{A}{4}$ and temperature $T=\frac{1}{2 \pi \tilde{r}_A}$ then the application of the first law of thermodynamics
$-dE=T\ud S$ gives \cite{a4} us the Friedmann equations
\begin{equation}
\dot{H} - \frac{k}{a^2} = - \frac{8 \pi}{n-1} (\rho +p)
\end{equation}
and
\begin{equation}
\label{e9}
H^2 + \frac{k}{a^2} = \frac{16 \pi}{n(n-1)}\rho\,.
\end{equation}
Here $-\ud E$ is the amount of energy crossing the apparent horizon in an infinitesimal time interval, $A(=n\Omega_n \tilde{r}_A^{n-1})$ is the
area of the apparent horizon with $\Omega_n \big(= \frac{\pi^{\frac{n}{2}}}{\Gamma (\frac{n}{2}+1)} \big)$, the volume of an n-dimensional unit sphere.
$\rho$ and $p$ are the energy density and pressure of the perfect fluid respectively. Evaluation of equation (\ref{e9}) requires the conservation
or the continuity equation. Let us now consider that the apparent horizon absorbs or radiates a particle with energy $-\ud E$. We can identify
this energy with the particle momentum uncertainty. If we apply the Heisenberg uncertainty principle to the situation then the increase or decrease
in area of the apparent horizon is given by
\begin{align}
\ud A &= \frac{4}{T}\ud E \nonumber \\
   &\approx \frac{4}{T}\frac{1}{\delta x}.
\end{align} 
If we consider the effect of GUP (equation (\ref{u3})) then the above equations should be modified. We now write
\begin{align}
\label{e11}
\ud A &\approx \frac{4}{T} \frac{1}{\delta x} f_{GUP}(\delta x^2) \nonumber \\
   &\approx \frac{4}{T} f_{GUP}(\delta x^2) \ud A\,.
\end{align}
Now $\delta x \approx 2 \tilde{r}_A = 2 \big(\tfrac{A}{n\Omega_n}\big)^{\frac{1}{n-1}}$. So $f_{GUP}(\delta x^2)$ can be written as a function of
$A$. We will now use $f_{GUP}(A)$ instead of $f_{GUP}(\delta x^2)$. A straightforward calculation gives the form of $f_{GUP}(A)$ in series form and
we write
\begin{eqnarray}
\label{f1}
\nonumber \\ &&
f_{GUP}(A)=  1 + \frac{\beta}{2}\,l_p\, \bigg(\frac{n\Omega_n}{A}\bigg)^{\frac{1}{n-1}} \nonumber \\
 && + \frac{\alpha^2 + \beta^2}{4}\,l_p^2 \,\bigg(\frac{n\Omega_n}{A}\bigg)^{\frac{2}{n-1}}  \nonumber \\
 &&+ \frac{3\alpha^2 \beta + \beta^3}{8} \,l_p^3\, \bigg(\frac{n\Omega_n}{A}\bigg)^{\frac{3}{n-1}}  \nonumber \\
 &&+ \frac{2\alpha^4 + \beta^4 + 6\alpha^2
\beta^2}{16}\,l_p^4\, \bigg(\frac{n\Omega_n}{A}\bigg)^{\frac{4}{n-1}} \nonumber \\
&&+ \frac{10\alpha^4 \beta + 10\alpha^2 \beta^3 + \beta^5}{32}\,l_p^5\, \bigg(\frac{n\Omega_n}{A}\bigg)^{\frac{5}{n-1}} \nonumber \\
&&+ \sum_{d=3} \bigg[ f_{2d}(\alpha \beta)\, l_p^{2d}\, \bigg(\frac{n\Omega_n}{A}\bigg)^{\frac{2d}{n-1}} \nonumber \\
&&+ f_{2d+1}(\alpha \beta)\,l_p^{2d+1}\, 
\bigg(\frac{n\Omega_n}{A}\bigg)^{\frac{2d+1}{n-1}}\,\,\bigg]\,\,.
\end{eqnarray}
Here $f_{2d}(\alpha \beta)$ and $f_{2d+1}(\alpha \beta)$ are polynomial functions of $\alpha$ and $\beta$. We now investigate our FRW universe
in (3+1)-dimension. By putting $n=3$ in equation (\ref{f1}) we get
\begin{eqnarray}
\label{f2}
\nonumber \\ &&
f_{GUP}(A) =1 + \frac{\beta}{2}\,l_p\, \bigg(\frac{4\pi}{A}\bigg)^{\frac{1}{2}} \nonumber \\ 
&&+ (\alpha^2 + \beta^2)\,l_p^2 \,\bigg(\frac{\pi}{A}\bigg) \nonumber \\
&&+ \frac{3\alpha^2 \beta + \beta^3}{8} \,l_p^3\, \bigg(\frac{4\pi}{A}\bigg)^{\frac{3}{2}} \nonumber \\
&&+ (2\alpha^4 + \beta^4 + 6\alpha^2 \beta^2)\,l_p^4\, \bigg(\frac{\pi}{A}\bigg)^2 \nonumber \\
&&+ \frac{10\alpha^4 \beta + 10\alpha^2 \beta^3 + 
\beta^5}{32}\,l_p^5\, \bigg(\frac{4\pi}{A}\bigg)^{\frac{5}{2}} \nonumber \\
&&+ \sum_{d=3} \bigg[ f_{2d}(\alpha \beta)\, l_p^{2d}\, \bigg(\frac{4\pi}{A}\bigg)^d \nonumber \\
&&+ f_{2d+1}(\alpha \beta)\,l_p^{2d+1}\, 
\bigg(\frac{4\pi}{A}\bigg)^{\frac{2d+1}{2}} \,\,\bigg]\,\,.
\end{eqnarray}
Using (\ref{f2}) with (\ref{e11}) we get the area with corrections as
\begin{eqnarray}
\label{ar}
\nonumber \\ &&
A_{GUP} =  A + (4\pi)^{\frac{1}{2}}\,\beta \,l_p\,\sqrt{A} \nonumber \\ 
&&+ \pi (\alpha^2+\beta^2)\,l_p^2\, \ln A  \nonumber \\
&&- (4\pi)^{\frac{3}{2}} \frac{(3\alpha^2 \beta + \beta^3)}{4} \,l_p^3\, A^{-\frac{1}{2}} \nonumber \\
&& - \pi^2 (2\alpha^4 + \beta^4 + 6\alpha^2 \beta^2) \, l_p^4\, A^{-1} \nonumber \\
&& - (4\pi)^{\frac{5}{2}} \frac{(10\alpha^4 \beta + 10 \alpha^2 \beta^3 + 
\beta^5)}{48}\,l_p^5\, A^{-\frac{3}{2}} \nonumber \\ 
&& - \sum_{d=3} \bigg[ f_{2d}(\alpha \beta)\, l_p^{2d}\, \frac{(4\pi)^d}{d-1} A^{1-d} \nonumber \\ 
&&+ 2f_{2d+1}(\alpha \beta)\,l_p^{2d+1}\, 
\frac{(4\pi)^{\frac{2d+1}{2}}}{2d-1}A^{\frac{1-2d}{2}} \,\,\bigg] \nonumber \\
&&+ C' \,\,,
\end{eqnarray}
where $C'$ is the integration constant. Following 
Bekenstein Hawking argument for the entropy area relation we
can calculate the GUP modified
entropy in this case. The entropy is given by
\begin{eqnarray}
\label{en}
\nonumber \\ &&
S_{GUP} = \frac{A}{4} + \pi^{\frac{1}{2}}\,\beta \,l_p\,\sqrt{\frac{A}{4}} \nonumber \\ 
&&+ \frac{\pi}{4} (\alpha^2+\beta^2)\,l_p^2\, \ln{\frac{A}{4}} \nonumber \\
&&- \pi^{\frac{3}{2}} \frac{(3\alpha^2 \beta + \beta^3)}{4} \,l_p^3\, \bigg(\frac{A}{4}\bigg)^{-\frac{1}{2}}  \nonumber \\
&& - \Big(\frac{\pi}{4}\Big)^2 (2\alpha^4 + \beta^4 + 6\alpha^2 \beta^2) \, l_p^4\, \bigg(\frac{A}{4}\bigg)^{-1} \nonumber \\
 &&- \pi^{\frac{5}{2}} \frac{(10\alpha^4 \beta + 10 \alpha^2 \beta^3 + \beta^5)}{48}\,l_p^5\, \bigg(\frac{A}{4}\bigg)^{-\frac{3}{2}} \nonumber \\
 &&- \sum_{d=3} \Bigg[ f_{2d}(\alpha \beta)\, l_p^{2d}\, \frac{\pi^d}{d-1} \bigg(\frac{A}{4}\bigg)^{1-d}  \nonumber \\ 
 &&+ 2f_{2d+1}(\alpha \beta)\,l_p^{2d+1}\, 
\frac{\pi^{\frac{2d+1}{2}}}{2d-1}\bigg(\frac{A}{4}\bigg)^{\frac{1-2d}{2}} \,\,\Bigg] \nonumber \\
&&+ C \,\,.
\end{eqnarray}
Here we see that a new correction $\sim \sqrt{A}$ is added to the entropy. This is a consequence of the linear term in Plank length in the
uncertainty relation. This new entropy bound different from the conventional ones was first pointed out in \cite{ali}. Here we would like
to avoid going into the debate about the sign of the prefactor of the correction terms. Considering $S(A)$ as the entropy of the apparent horizon
we can obtain Friedmann equations by applying the first law of thermodynamics \cite{a13},\cite{a}. The equations are 
\begin{equation}
\label{fd1}
\Big(\dot{H} - \frac{k}{a}\Big) S'(A) = -\pi (\rho + p)
\end{equation}
and
\begin{equation}
\label{fd2}
\frac{8\pi \rho}{3} = - \pi \int S'(A) \Big(\frac{A}{4}\Big)^{-2} \ud A.
\end{equation}
The prime denotes the derivative with respect to $A$. We will consider $S_{GUP}(A)$ as the entropy of the apparent horizon due to the GUP
considered. Using (\ref{en}), (\ref{fd1}) and (\ref{fd2}) we finally get the modified Friedmann equations as 
\begin{eqnarray}
\label{mod1}
\nonumber \\ &&
-4\pi(\rho + p) = \Big(\dot{H} -\frac{k}{a}\Big) \Bigg[ 1 + \frac{\beta}{2}\,l_p\, \bigg(\frac{4\pi}{A}\bigg)^{\frac{1}{2}} \nonumber \\ &&
 + (\alpha^2 + \beta^2)\,l_p^2 \,
\bigg(\frac{\pi}{A}\bigg)  \nonumber \\
&& + \frac{3\alpha^2 \beta + \beta^3}{8} \,l_p^3\, \bigg(\frac{4\pi}{A}\bigg)^{\frac{3}{2}} \nonumber \\ &&
 + (2\alpha^4 + \beta^4 + 6\alpha^2 \beta^2)\,l_p^4\, \bigg(\frac{\pi}{A}\bigg)^2   \nonumber \\
&& + \frac{10\alpha^4 \beta + 10\alpha^2 \beta^3 + \beta^5}{32}\,l_p^5\, \bigg(\frac{4\pi}{A}\bigg)^{\frac{5}{2}} \nonumber \\ &&
 + \sum_{d=3} \bigg\{ f_{2d}(\alpha \beta)\, l_p^{2d}\, \bigg(\frac{4\pi}{A}\bigg)^d \nonumber \\ &&
 + f_{2d+1}(\alpha \beta)\,l_p^{2d+1}\, 
\bigg(\frac{4\pi}{A}\bigg)^{\frac{2d+1}{2}} \,\,\bigg\}\Bigg] 
\end{eqnarray}
and
\begin{eqnarray}
\label{mod2}
\nonumber \\ &&
\frac{8\pi \rho}{3} = \, 4\pi \Bigg[ \frac{1}{A}
+ (4\pi)^{\frac{1}{2}} \frac{\beta \, l_p\,}{3} A^{-\frac{3}{2}}  \nonumber \\ && 
+ \pi \frac{\alpha^2 +\beta^2}{2}\,l_p^2\, A^{-2} \nonumber \\ &&
+ (4\pi)^{\frac{3}{2}} \frac{3\alpha^2 \beta + \beta^3}{20}\,l_p^3\, A^{-\frac{5}{2}}  \nonumber \\ &&
 + \pi^2 \frac{2\alpha^4 + \beta^4 + 6\alpha^2 \beta^2}{3} \,l_p^4\,A^{-3}  \nonumber \\ &&
+ (4\pi)^{\frac{5}{2}} \frac{10\alpha^4 \beta + 10 \alpha^2 \beta^3 + \beta^5}{112} \,l_p^5\, A^{-\frac{7}{2}} \nonumber \\ &&
+ \sum_{d=3} \bigg\{ f_{2d}(\alpha \beta)\, l_p^{2d}\, \frac{(4\pi)^d}{d+1} A^{-d-1} \nonumber \\ &&
 + 2f_{2d+1}(\alpha \beta)\,l_p^{2d+1}\, 
\frac{(4\pi)^{\frac{2d+1}{2}}}{2d+3}A^{\frac{-2d-3}{2}} \,\,\bigg\}\Bigg] \, \, .
\end{eqnarray}
Now if we set $\beta=0$ in equations (\ref{mod1}) and (\ref{mod2}) we get back equations (26) and (27) of \cite{a} where the uncertainty relation
was considered to be $\delta x \delta p \geq 1 + \alpha^2\,l_p^2\,(\delta p)^2$. Now we can easily tune the parameters $\beta=-2a_0$ and $\alpha^2=4a_0^2$ so
that they satisfy equation (\ref{ee1}). Interestingly the dominant correction term in the GUP modified entropy is the $\sqrt{A}$ term. We also
see that there are new corrections with fractional power of $A$ like $A^{-\frac{1}{2}}$, $A^{-\frac{3}{2}}$ and so on. This is quite different from
our conventional knowledge. We know that GUP effects the dynamics of the black holes and our early universe where very high energy effects seem important.
In this article we have derived the modified Friedmann equations considering the Generalized Uncertainty Principle mentioned earlier. In the
midway we found new correction terms being contributed to the existing entropy-area relation. We hope that our investigation might
be helpful in providing a better understanding of the physics in the Plank regime. 

\acknowledgments
The author is very thankful to Prof. Narayan Banerjee for helpful guidance and enlightening discussions. The author would also like to
thank an anonymous referee for useful comments and suggestions.

\makeatletter
\let\clear@thebibliography@page=\relax
\makeatother

\end{document}